\documentclass[onecolumn]{article}          
\usepackage[]{graphicx}
\usepackage{amsmath}
\usepackage{amssymb}
\usepackage[]{graphicx}
\usepackage{epsfig}
\usepackage{color}
\usepackage{subfigure}
\usepackage{epstopdf}

\def\les{\left[}
\def\ris{\right]}

\def\##1{{\bf #1}}
\def\=#1{\underline{\underline #1}}

\def\ux{\hat{\#u}_{\rm x}}
\def\uy{\hat{\#u}_{\rm y}}
\def\uz{\hat{\#u}_{\rm z}}
\def\uprop{\hat{\#u}_{\rm prop}}
 
\def\eps{\varepsilon}

\def\epso{\eps_{\scriptscriptstyle 0}}
\def\lambdao{\lambda_{\scriptscriptstyle 0}}
\def\muo{\mu_{\scriptscriptstyle 0}}
\def\ko{k_{\scriptscriptstyle 0}}

\def\nmet{n_{\rm met}}
\def\Lmet{L_{\rm met}}

\def\ffl{f_{\ell}}
\def\nfl{n_{\ell}}

\def\fnp{f_{\rm np}}
\def\nnp{n_{\rm np}}

\def\thetainc{\theta_{\rm inc}}
\def\thetaSPP{\theta_{\rm SPP}}

\def\nprism{n_{\rm prism}}

\def\vph{v_{\rm ph}}
\def\propdist{\Delta_{\rm prop}}

\def\Rp{R_{\rm p}}
\def\Rs{R_{\rm s}}

\begin{document}

\title{Enhancement of dynamic sensitivity of multiple surface-plasmonic-polaritonic sensor using silver nanoparticles}


\author{Farhat Abbas, Muhammad Faryad\\
Department of Physics, Lahore University of Management Sciences,\\ Lahore, 54792, Pakistan. Email: muhammad.faryad@lums.edu.pk\\\\
Stephen E. Swiontek, Akhlesh Lakhtakia\\
              Department of Engineering Science and Mechanics, Pennsylvania\\ State University, University Park, Pennsylvania 16802, USA. \\
Email: akhlesh@psu.edu
}

\maketitle
\begin{abstract}
Multiple surface plasmon-polariton (SPP)  waves  excited at the interface of a homogeneous isotropic metal and a chiral sculptured thin film (STF)  impregnated with silver nanoparticles were theoretically assessed for the multiple-SPP-waves-based sensing of a fluid uniformly infiltrating the chiral STF. The Bruggemann homogenization formalism was used in two different modalities to determine the three principal relative permittivity scalars of the silver-nanoparticle-impregnated chiral STF infiltrated uniformly by the fluid. 
 The  dynamic sensitivity  increased when silver nanoparticles were present, provided their volume fraction did not exceed about 1\%.  
\end{abstract}

\maketitle


\section{Introduction}
Optical sensors that rely on the variability of the excitation of a surface plasmon-polariton (SPP) wave guided by the planar interface of a metal and a homogeneous isotropic dielectric (HID) material with respect to the refractive index of the latter material are widely used to detect and
quantify the concentrations of analytes \cite{Homola2003,AZL,CZM}. At any given free-space wavelength $\lambdao$,
only one  SPP wave can be excited in a sensor of this kind  \cite{M2007} and 
 can be used to detect a single
analyte present in the HID material \cite{AZL,CZM}. 

Despite
successes,  strategies to economically  \cite{LMG} improve reliability    continue
to be devised. One   strategy is to disperse a small volume fraction of metal nanoparticles in the HID material \cite{EH2001,HWL,Bedford},
which is usually a fluid for sensing applications. This strategy exploits the excitation of the 
Fr\"ohlich mode \cite{BHbook}
or a localized surface-plasmon resonance \cite{Kalele}
in a metal nanoparticle by an SPP wave.
Metal nanoparticles have been reported to increase the sensitivity, depending on the substrate as well as the shape and material of the nanoparticles~\cite{EH2001,HWL,KSL,LIAL}.   Also,  metal nanoparticles are necessary to provide sites for recognition molecules that would bind to the specific analyte
 to be detected \cite{Saha}.  
  
 An emerging strategy is to replace the HID material by a periodically nonhomogeneous dielectric (PHD) material  that is porous \cite{SPL2013}. The planar
 interface of a metal and a  PHD material is capable  of guiding multiple SPP waves of the same frequency and direction of propagation but with different polarization states, phase speeds, spatial profiles, attenuation rates, and degrees of localization to the interface \cite{PML2013}. If the PHD material is porous, it can be infiltrated with the analyte to be sensed.  Multiple SPP waves provide the opportunity for more reliable and more sensitive optical sensing than  a single SPP wave. This has been established both theoretically and experimentally  using   sculptured thin films (STFs) as  porous PHD materials \cite{SPL2013,ML2012,FA2015}. However, the effect of infiltration
 of the STF by metal nanoparticles in a
 multiple-SPP-waves-based sensing scenario is not known.

Therefore, we set out to investigate  multiple-SPP-waves-based sensing in which  a chiral STF \cite{LM2005}
is infiltrated by metal nanoparticles as well as a fluid. A chiral STF is an ensemble of parallel  nanohelixes,
and is therefore macroscopically conceptualized as an anisotropic and helically nonhomogeneous material in the optical spectral regime.
We first determined the complex wave\-numbers $q$
of SPP waves in a canonical boundary-value problem in which a metal and a chiral STF occupy adjoining half spaces.  
Then we determined the angle of incidence $\thetainc$ of light that is needed to excite an SPP
wave in the Turbadar--Kretschmann--Raether prism-cou\-pl\-ed
configuration \cite{PML2013} employed for experiments.

The plan of this paper is as follows: Sec. \ref{mod} describes the calculation of  three nanoscale parameters by using the inverse Bruggeman homogenization formalism and the use of the forward Bruggeman homogenization formalism to determine the constitutive parameters of the chiral STF \cite{ML2012} in which silver nanoparticles are placed and whose void regions are in infiltrated by a fluid. In Sec. \ref{nrd}, numerical results of the canonical 
boundary-value problem and the prism-coupled configuration are presented and discussed. Concluding remarks are presented in Sec. \ref{con}. 
An exp($-i\omega t$) dependence on time $t$ is implicit, where $\omega$ is the angular frequency and $i=\sqrt{-1}$. The Cartesian unit vectors are denoted by $\ux$, $\uy$, and $\uz$; dyadics are double underlined; vectors are in bold face; and the free-space wavenumber and wavelength are denoted by $\ko=\omega \sqrt{\epso \muo}$ and $\lambdao=2\pi/\ko$, respectively, where $\muo$ is the permeability of free space and $\epso$ is the permittivity of free space.

\section{Theory in Brief}
\subsection{Microscopic-to-continuum model for chiral STFs}
\label{mod}
Chiral STFs are usually fabricated using physical vapor deposition  by directing a vapor flux towards a rotating substrate at an angle $\chi_{\rm v}$
\cite{LM2005}.  The angle $\chi_{\rm v}$, the deposition rate, and the rotation speed of the substrate are the main factors on which the porosity  of the chiral STF depends. The relative permittivity dyadic of a chiral STF is stated as 
\begin{equation}
\=\eps_{Chi\,STF}(z) = \=S_{z}(z)\cdot \=S_{y}(\chi) \cdot   \=\eps^\circ_{\rm ref} \cdot \=S^{-1}_{y}(\chi) \cdot \=S^{-1}_{z}(z)\,,\label{epssntf}
\end{equation}
with the dyadics 
\begin{equation}
\left. \begin{aligned}
\=S_{z}(z) &= \uz \uz + ( \ux \ux + \uy \uy)\cos(h \pi z/\Omega)\\& + (\uy \ux - \ux \uy)\sin(h \pi z/\Omega) \\
\=S_{y}(\chi) &= \uy \uy + ( \ux \ux + \uz \uz)\cos\chi\\& + (\uz \ux - \ux \uz)\sin\chi \\
\=\eps^\circ_{\rm ref} &= \eps_a \uz \uz + \eps_b \ux \ux + \eps_c \uy \uy
\end{aligned}
\right \}\,;
\end{equation}
where the direction of helical nonhomogeneity is parallel to the $z$ axis, 
$\chi\in(0,\pi/2]$ is the angle of
rise of the nanohelixes in the chiral STF with respect to the $xy$ plane, $2\Omega$ is the  helical period, and $h=+1$ for structural right handedness but $h=-1$ for structural left handedness.

The three principal relative permittivity scalars $\eps_a$, $\eps_b$, and $\eps_c$ will be different for an as-deposited chiral STF than for a chiral STF
containing the metal nanoparticles and/or infiltrated by
a fluid. The process to determine these quantities for the present work relies  on both experimental data and a theoretical 
microscopic-to-conti\-nuum model \cite{ML2012}. The as-deposited chiral STF is supposed to be made of a material
of refractive index $n_s$, this material assumed to be deposited in the form of   electrically
small ellipsoids  with a transverse aspect ratio $\gamma_b$
somewhat larger than unity and a slenderness ratio $\gamma_\tau \gg 1$; i.e.,
every nanohelix is a string of ellipsoidal sausages. The volume fraction of the
deposited material is denoted by $f_s\in[0,1]$. The void regions are  thought of as electrically small spheres of unit refractive index. 
The metal nanoparticles are
assumed to be electrically small spheres of refractive index $\nnp$.
The fluid
is taken to be distributed as electrically small spheres of refractive index $\nfl$.

As discussed elsewhere \cite{ML2012}, $h$ is fixed during
deposition, while  $\Omega$ and $\chi$ can be determined from scanning electron micrographs
of the as-deposited chiral STF. Let $\eps_a=\eps_{a1}$, $\eps_b=\eps_{b1}$, and $\eps_c=\eps_{c1}$
for the as-deposited chiral STF. With the assumptions that (i) $\eps_{a1,b1,c1}$ have been measured through
optical experiments, (ii) $\gamma_\tau = 15$, and (iii) the as-deposited
chiral STF is a composite material comprising the deposited material and
air, the inverse Bruggeman formalism can be employed to determine $n_s$, $f_s$, and $\gamma_b$ \cite{ML2012}. Let us
note that the effect of increasing   $\gamma_{\tau}$  beyond $10$ is insignificant \cite{PML2013}.

Suppose next that $\eps_a=\eps_{a2}$, $\eps_b=\eps_{b2}$, and $\eps_c=\eps_{c2}$ for a chiral STF uniformly infiltrated by metal 
nano\-spheres and/or fluid  nano\-spheres. The volume fraction of the metal is denoted by $\fnp\in[0,1]$, while that
of the fluid is denoted by $\ffl=1-\fnp-f_s\in[0,1]$. With $n_s$, $f_s$,  $\gamma_b$, and $\fnp$ known, 
$\eps_{a2,b2,c2}$ can be determined using the forward Bruggeman formalism \cite{AL2007}. Obviously, $\fnp=0$
if there are no metal nanoparticles \cite{PML2013,ML2012}. Furthermore, if the fluid is absent, we have to set
$\nfl = 1$ while implementing the forward Bruggeman formalism.

\subsection{Canonical boundary-value problem}
\begin{figure}[!ht]
\begin{center}
\subfigure[]{\includegraphics[width=0.45\textwidth]{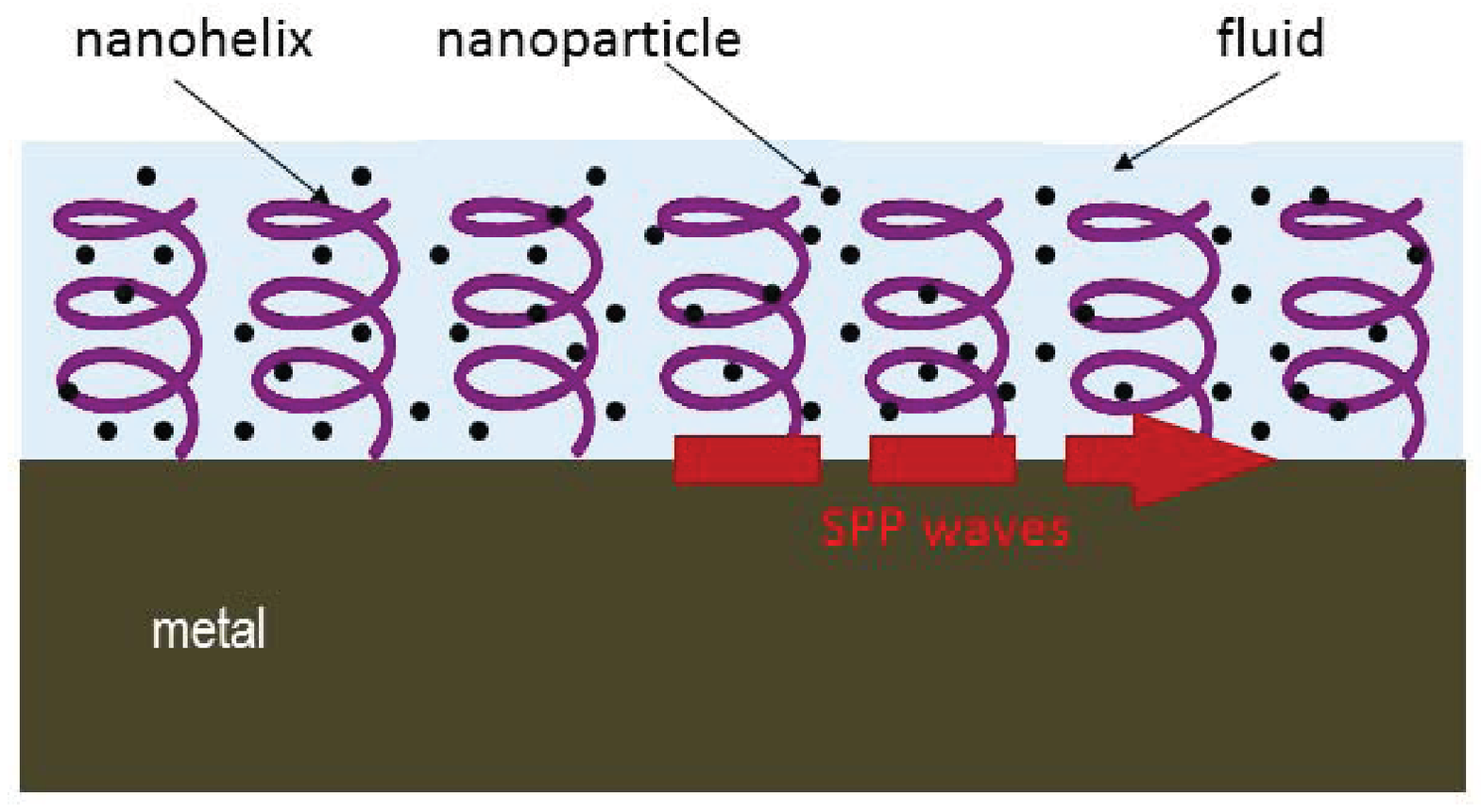}}
\subfigure[]{\includegraphics[width=0.45\textwidth]{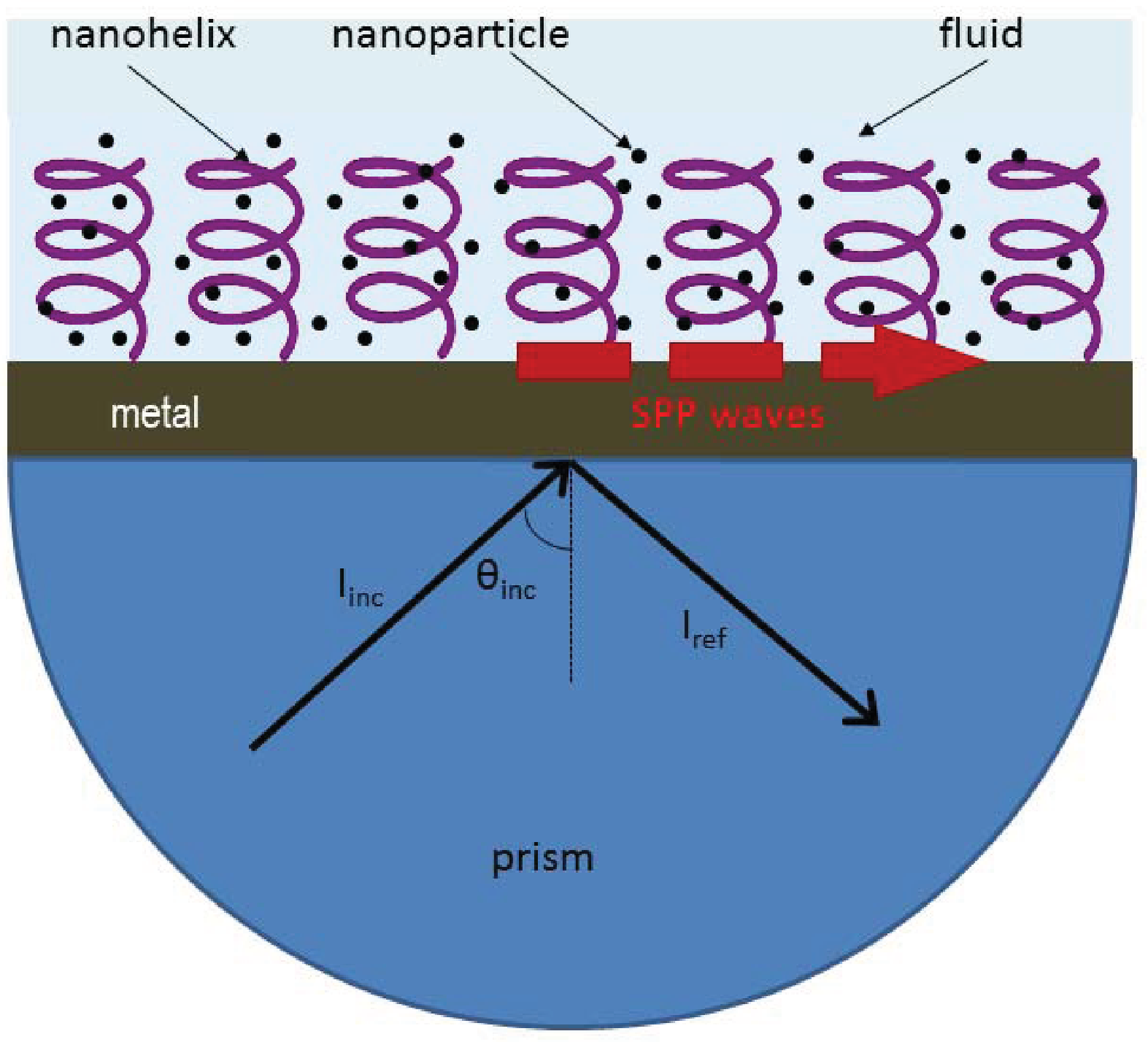}}
\caption{Schematics of  (a) the canonical boundary-value problem, and (b)
the Turbadar--Kretschmann--Raether prism-coupled configuration. 
} \label{Fig1}
\end{center}
\end{figure}

In order to formulate the canonical problem for SPP wave propagation \cite{PML2013}, we take the half space $z<0$ to be wholly
occupied by a metal
of refractive index $\nmet$ and the half space $z >0$ by the chiral STF, as shown in Fig.~\ref{Fig1}(a).

The SPP wave is taken to propagate parallel to the unit vector 
$\uprop=\ux\cos\psi+\uy\sin\psi$,
$\psi\in[0,2\pi)$, in the $xy$ plane  and to decay far away from the interface $z=0$. The electric and magnetic field phasors can be written everywhere as
\begin{eqnarray}
\nonumber
&&
\left.\begin{array}{l}
\#E(x,y,z)= \#e(z) \exp[{iq}(x\cos\psi+ y\sin\psi)]\\[5pt]
\#H(x,y,z)= \#h(z)\exp[{iq}(x\cos\psi+ y\sin\psi)]
\end{array}\right\}\,, 
\\
&&\qquad\qquad\quad z \in(-\infty,\infty)\,,
\label{eq:EH1}
\end{eqnarray}
where the complex-valued wavenumber $q$ and the vector functions
$\#e(z)$ and $\#h(z)$ are not known.  The procedure to obtain a dispersion equation for  $q$
is provided in detail elsewhere \cite[Chap.~3]{PML2013}, and so is the procedure to determine the functions $\#e(z)$ and $\#h(z)$
  for each solution $q$ of the dispersion equation.

\subsection{Prism-coupled configuration}
The prism-coupled configuration shown in Fig.~\ref{Fig1}(b) is practically implementable. Light of fixed
polarization state and free-space wavelength $\lambdao$ is incident
on the semicircular face of the prism at an angle $\thetainc$, the refractive index of the
prism being denoted by $\nprism$. The base of the prism is coated with a
metal  layer of thickness $\Lmet$ and refractive index $\nmet$. The other face of metal layer is in intimate contact
with a chiral STF of thickness $L_d$, beyond which is a half space occupied by the fluid. With $I_{\rm inc}$ as the
intensity of light inside the prism impinging upon the prism/metal interface and $I_{\rm ref}$ as the
intensity of light inside the prism reflected by the prism/metal interface, the detailed procedure to determine the ratio
$R=I_{\rm ref}/I_{\rm inc}$ is available elsewhere \cite{PML2013}.

The minimums  of $R$ are identified
as the angle $\thetainc\in[0,\pi/2)$ is  varied.
If a certain minimum occurs at the same value $\thetaSPP$ of $\thetainc$ for all $L_d$ greater than
some threshold value while no transmission occurs in to the half space occupied by the
fluid, an SPP wave guided by the metal/chi\-ral-STF interface is said to be excited. Then,
\begin{equation}
\label{predict}
\thetaSPP=\sin^{-1}\les\frac{{\rm Re}\left(q\right) }{\ko\nprism}\ris\,
\end{equation}
ideally, where $q$ is a solution of the dispersion equation of the canonical boundary-value problem.

\section{Numerical Results and Discussion}
\label{nrd}
For all numerical results presented in this paper, we fixed $\lambdao=633$~nm
and set $\eps_{a1}=3.2589$, $\eps_{b1}=4.4571$, $\eps_{c1}=3.7829$, and
 $\chi=58.9928^\circ$, 
 in accord with previous works \cite{PML2013,ML2012,HWH1998}. The
inverse Bruggeman formalism yields $n_s=3.0517$, $f_s=0.5039$, and $\gamma_b=1.8381$ \cite{ML2012}.
We also chose $h=1$,   $\Omega=200$~nm, $\nprism=2.6$ (zinc selenide), 
$\nmet=1.3799 + 7.6095i$ (aluminum), $\Lmet=15$~nm, and $\nnp=0.0562+4.2776i$ (silver). Without significant loss
of generality, we fixed $\psi=0$. The parameters $L_d$, $\fnp$, and $\nfl\in[1,1.5]$ were
kept as variables. We confined the search for solutions to ${\rm Re}(q)/\ko\in[1,3]$.

\subsection{Chiral STF without  nanoparticles ($\fnp=0$)} \label{WSP}
In order to assess the effects of metal nanoparticles residing inside a chiral STF,
let us begin with the sensing of a fluid when $\fnp=0$ \cite{ML2012}.

\begin{figure}[!ht]
\begin{center}
\subfigure[]{\includegraphics[width=0.45\textwidth]{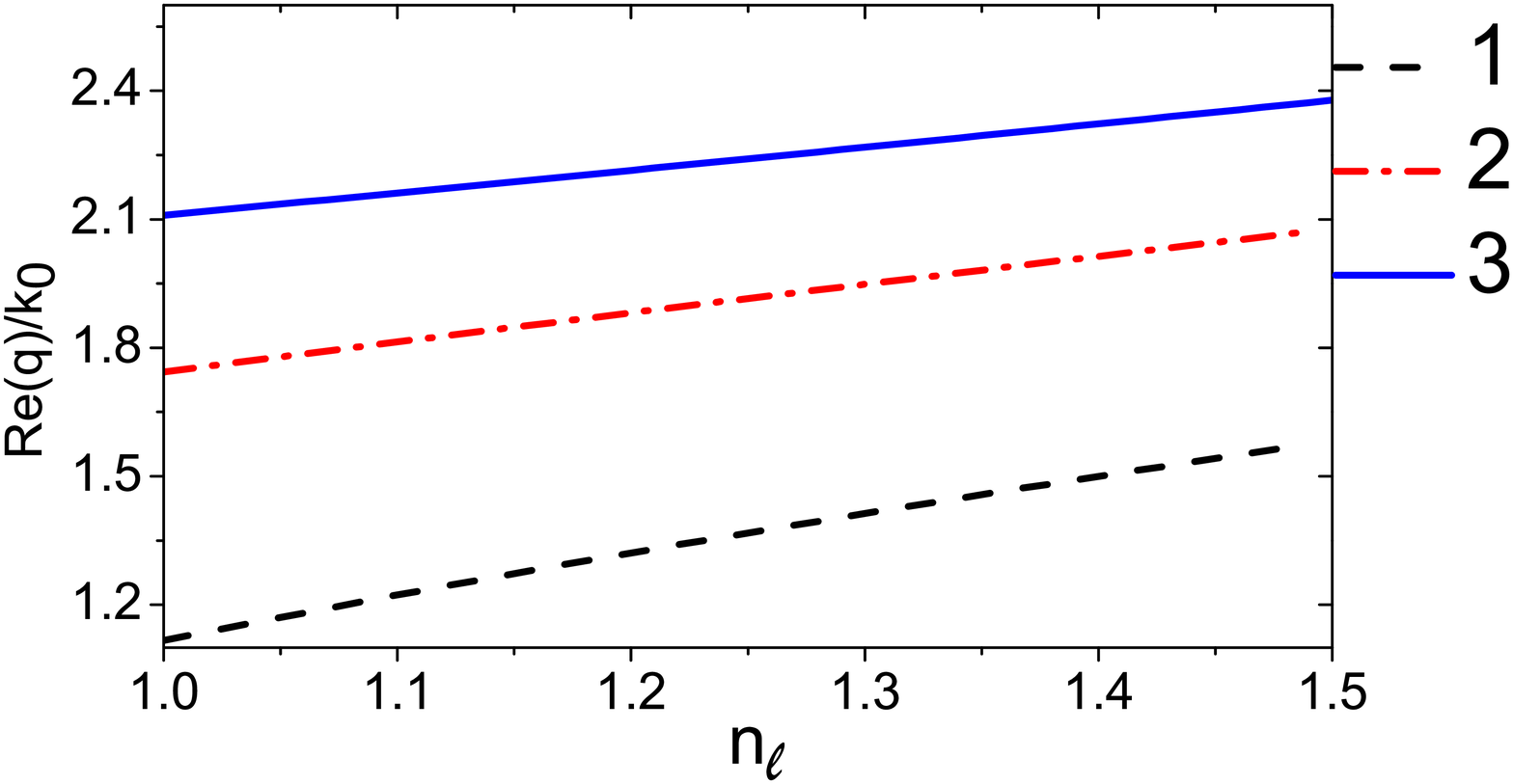}}
 \subfigure[]{\includegraphics[width=0.45\textwidth]{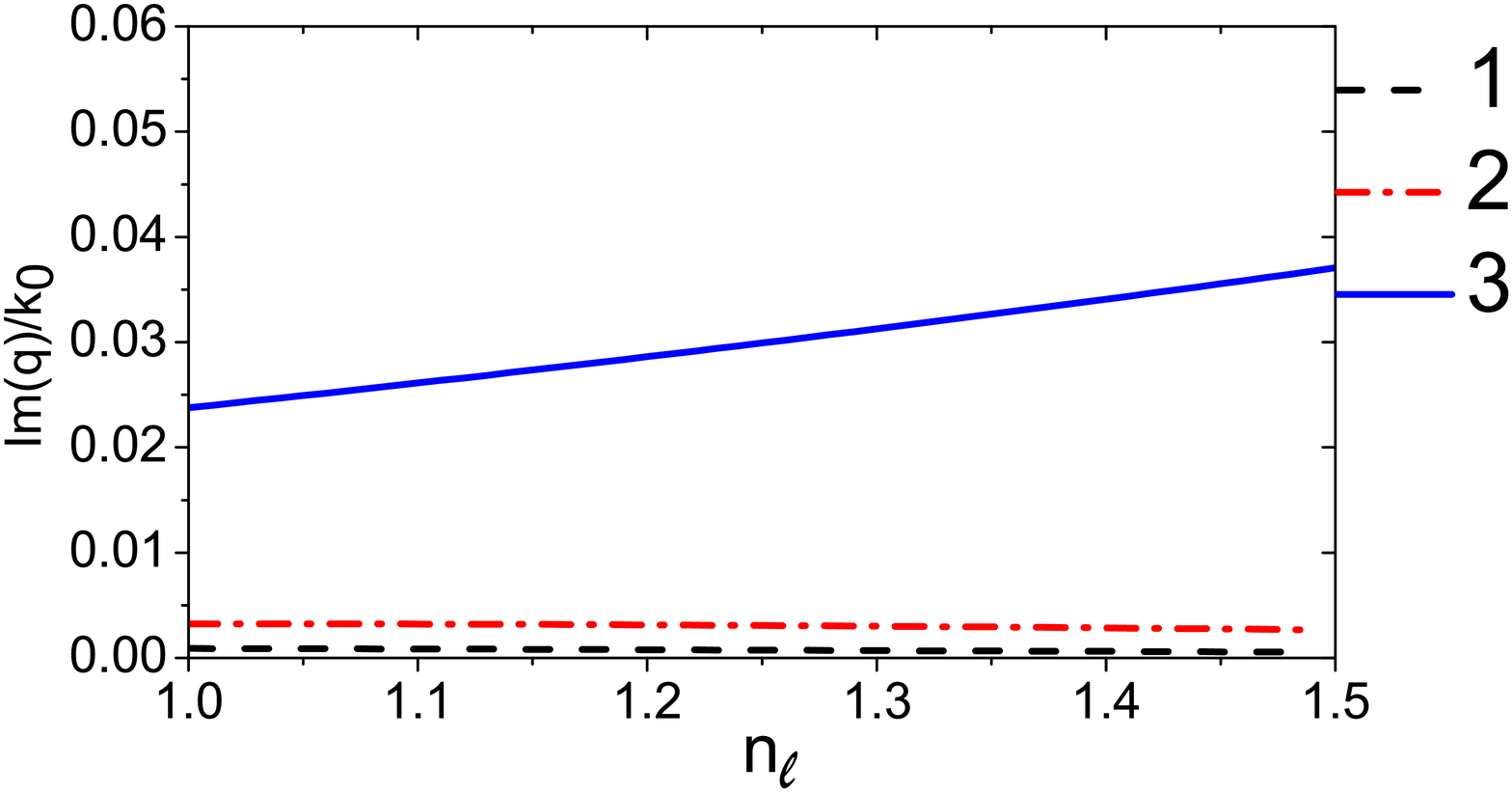}}
\caption{(a) Real and (b) imaginary parts of $q/\ko$ plotted against $\nfl$ when $\fnp=0$. The solutions
found were organized in three branches labeled
`1', `2', and `3'.
} \label{kappa1}
\end{center}
\end{figure}

\begin{figure}[!ht]
\begin{center}
\subfigure[]{\includegraphics[width=0.45\textwidth]{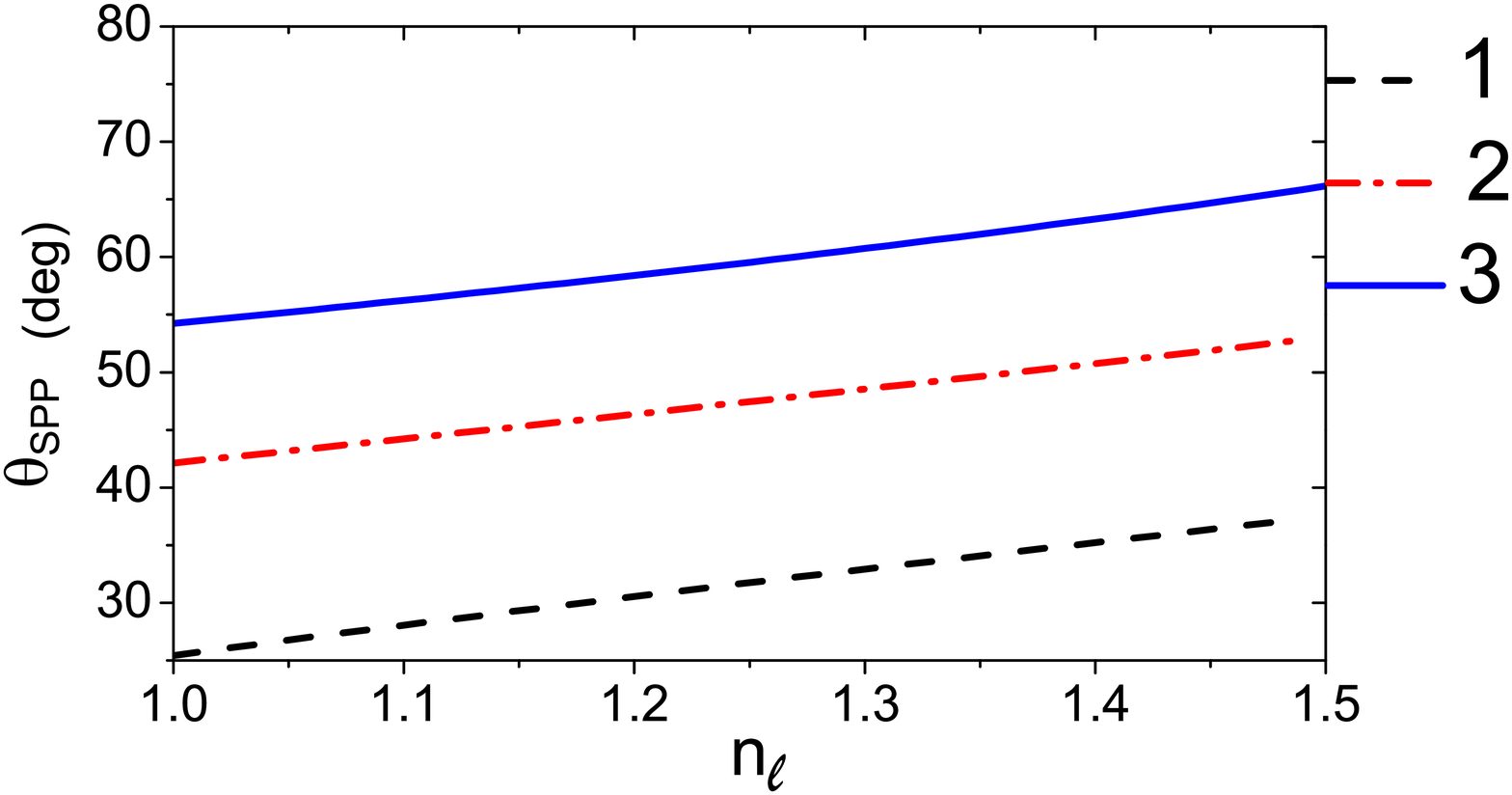}}
\subfigure[]{\includegraphics[width=0.45\textwidth]{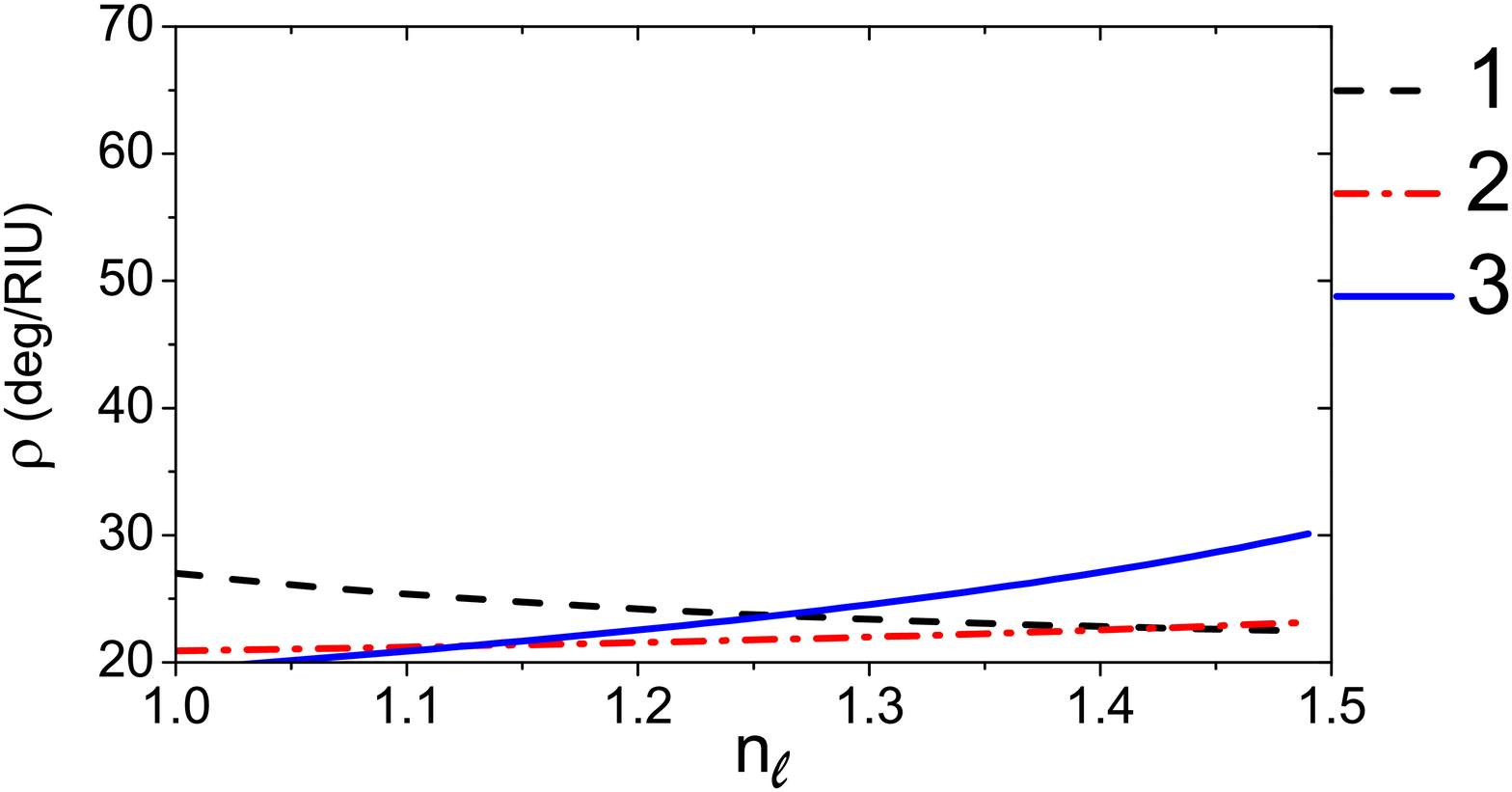}}
\caption{(a) $\thetaSPP$ predicted for the prism-coupled configuration
by using  the data of Fig.~\ref{kappa1} in Eq.~(\ref{predict}),
and (b) the  dynamic  sensitivity $\rho$ as a  function of $\nfl$ 
for each branch when $\fnp=0$. 
The acronym RIU stands for ``refractive-index unit".
}\label{thetarho1}
\end{center}
\end{figure}

The real and imaginary parts of the relative wave\-numbers $q/\ko$ are plotted in Fig. \ref{kappa1} as   functions of the refractive index 
$\nfl\in[1,1.5]$. All solutions of the dispersion equation for SPP waves are organized in three branches labeled
`1', `2', and `3'. The branches  differ in the ranges of both
the phase speed
$\vph=\omega/ {\rm Re} (q)$
and the propagation distance
$\propdist=1/{\rm Im}(q)$ in the $xy$ plane. Values of $\thetaSPP$ and the dynamic 
sensitivity \cite{Homola2003} 
\begin{equation}
\rho = \frac{d \thetaSPP(\nfl)}{d \nfl}\,
\label{ro}
\end{equation}
for all three branches are plotted against $\nfl$ in 
Fig.~\ref{thetarho1}, which shows that (i)  $\theta_{\rm inc}$ is very sensitive to changes in $\nfl$ and (ii)  $\rho$ can be as high as $30$~deg/RIU (branch 3).

\begin{figure}[!ht]
\begin{center}
\subfigure[]{\includegraphics[width=0.45\textwidth]{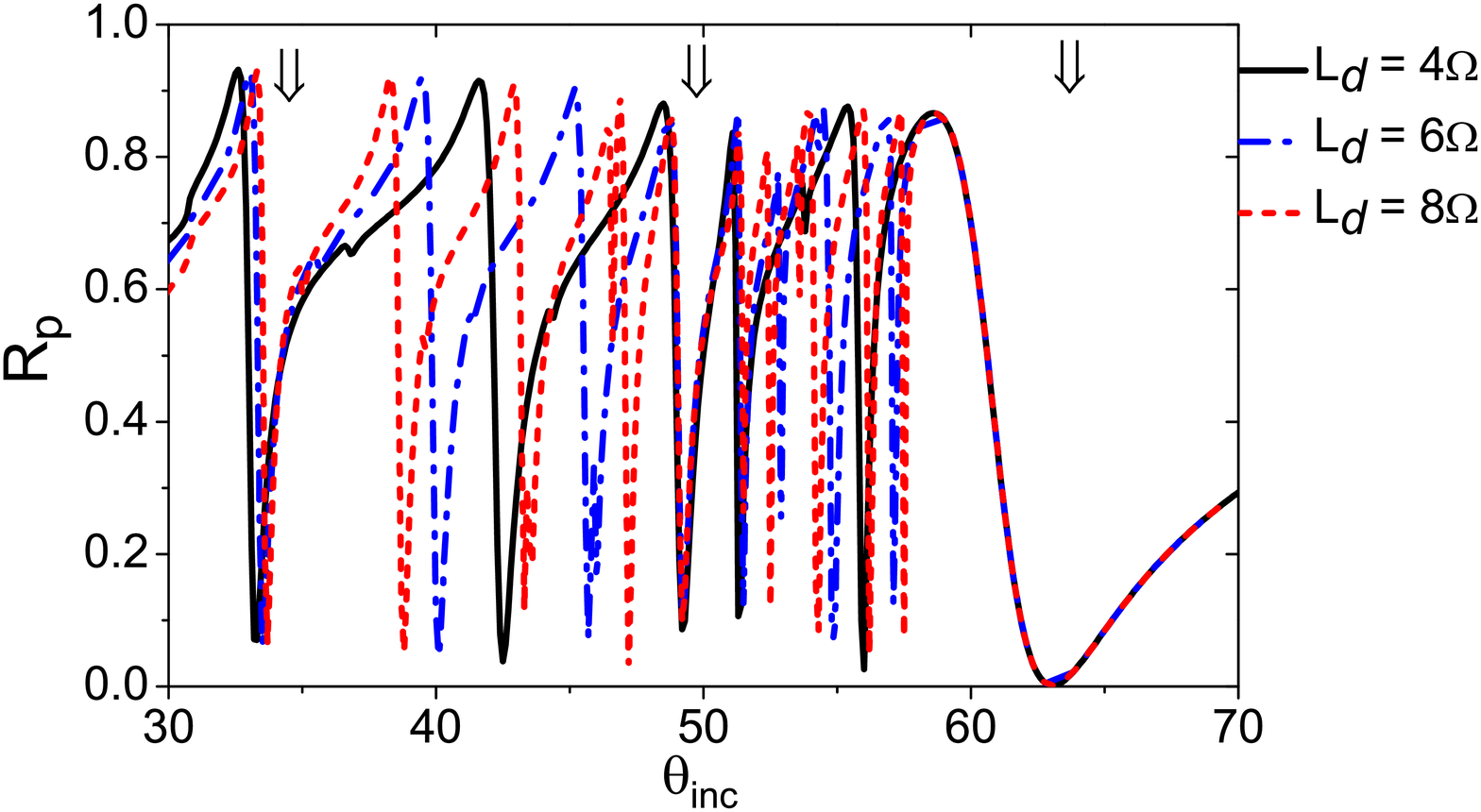}}
\subfigure[]{\includegraphics[width=0.45\textwidth]{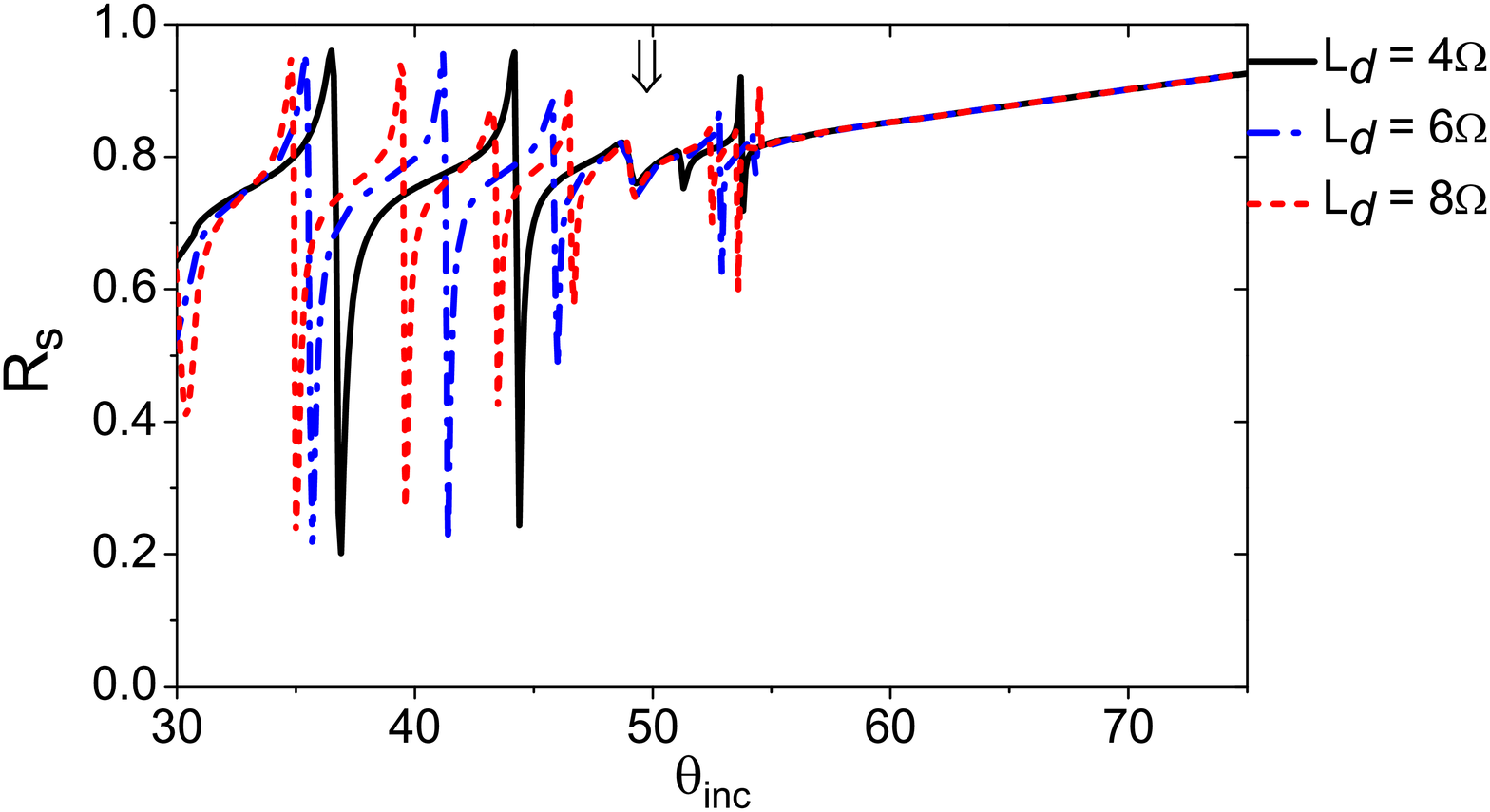}}
\caption{Reflectances (a) $\Rp$ and (b) $\Rs$ as  functions of $\thetainc$ for $L_d \in \{4\Omega, 6\Omega, 8\Omega \}$, 
when $\fnp=0$ and $\nfl=1.33$. The arrows indicate the minimums that represent the excitation of SPP waves. 
}\label{Rps_1}
\end{center}
\end{figure}

When  $\nfl=1.33$ (water), then SPP waves are predicted to be excited
at $\thetaSPP\in\left\{33.6^\circ,49.2^\circ,61.5^\circ\right\}$ in the prism-coupled
configuration, according to Fig.~\ref{thetarho1}(a). Figure~\ref{Rps_1}
 presents the reflectances $\Rp$ and $\Rs$ as  functions of $\theta_{\rm inc}$ for incident $p$- and $s$-polarized light,
 respectively, when  $L_d \in \{4\Omega, 6\Omega, 8\Omega \}$.  From~Fig.~\ref{Rps_1}(a), we conclude that
 SPP waves are excited  at $\thetainc\in \left\{33.5^\circ,\right.$ $\left.49.2^\circ,62.5^\circ\right\}$ because the angular locations of the minimums of $\Rp$
 are very weak\-ly dependent on $L_d$ when that thickness is large enough. We have also confirmed that these
 angular locations do not change if $\Lmet$ is increased from 15~nm to 20~nm.  Similarly, from Fig.~\ref{Rps_1}(b) we conclude that incident
 $s$-polarized light is able to excite an SPP wave at $\thetainc=49.3^\circ$. Thus, the predictions from the canonical
 boundary-value problem match the conclusions one can draw for the prism-coupled configuration.

\subsection{Chiral STF with silver nanoparticles ($\fnp>0$)}
Let us now present results for $\fnp>0$, i.e., when the chiral STF contains silver nanoparticles. In practice, $\fnp$ must be very small
or the chiral STF would become highly conducting, leading to the disappearance of SPP waves. For theoretical
work, $\fnp$ must be very small
or the Bruggeman formalism will yield unphysical results \cite{AL2007,Mackay2007}.

\begin{figure}[!ht]
\begin{center}
\subfigure[]{\includegraphics[width=0.45\textwidth]{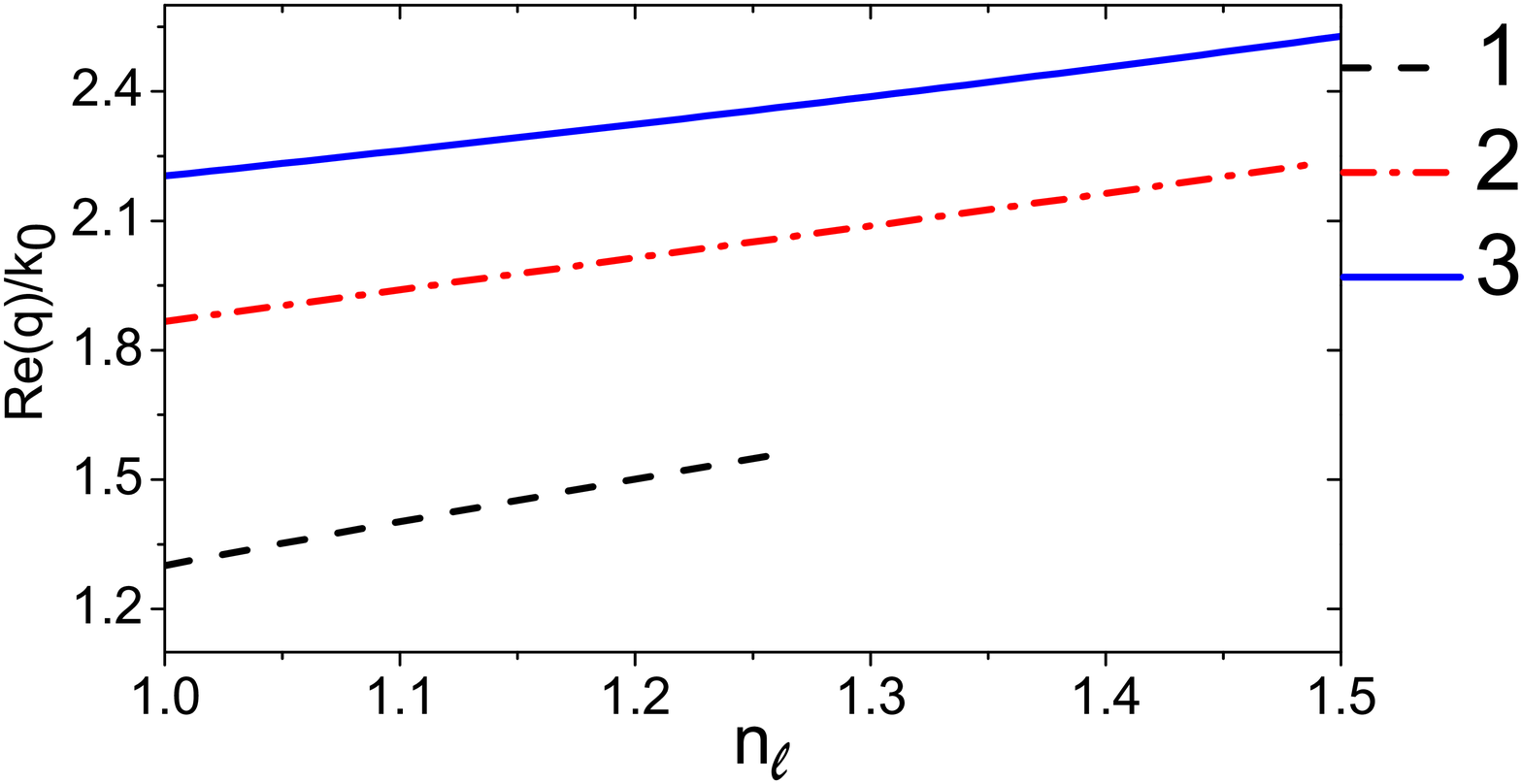}}
 \subfigure[]{\includegraphics[width=0.45\textwidth]{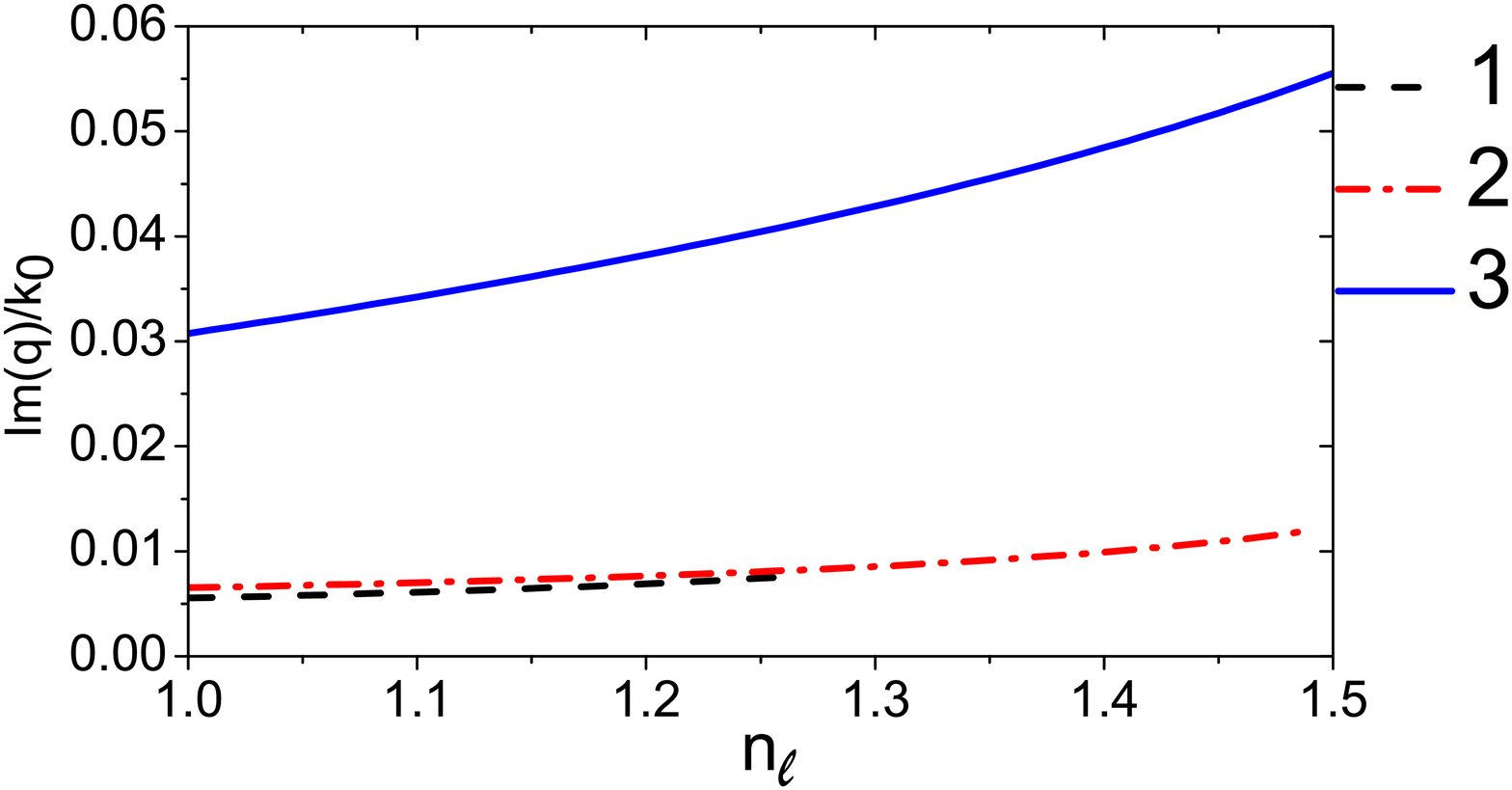}}
\caption{Same as Fig.~\ref{kappa1} except that the volume fraction of silver nanoparticles is $\fnp=0.01$. } \label{kappa2}
\end{center}
\end{figure}

Figure \ref{kappa2} presents the real and imaginary parts of $q/\ko$ when $\fnp = 0.01$. Just
as for $\fnp=0$ in Fig.~\ref{kappa1}, three branches of
SPP waves exist in Fig.~\ref{kappa2}. The presence of the silver nanoparticles tends to reduce
both the phase speed $\vph$ and the propagation distance $\propdist$.
Even more significantly, the higher-$\vph$ branches 1 and 2
 terminate at lower values of $\nfl$
when the silver nanoparticles are present.

\begin{figure}[!ht]
\begin{center}
\subfigure[]{\includegraphics[width=0.45\textwidth]{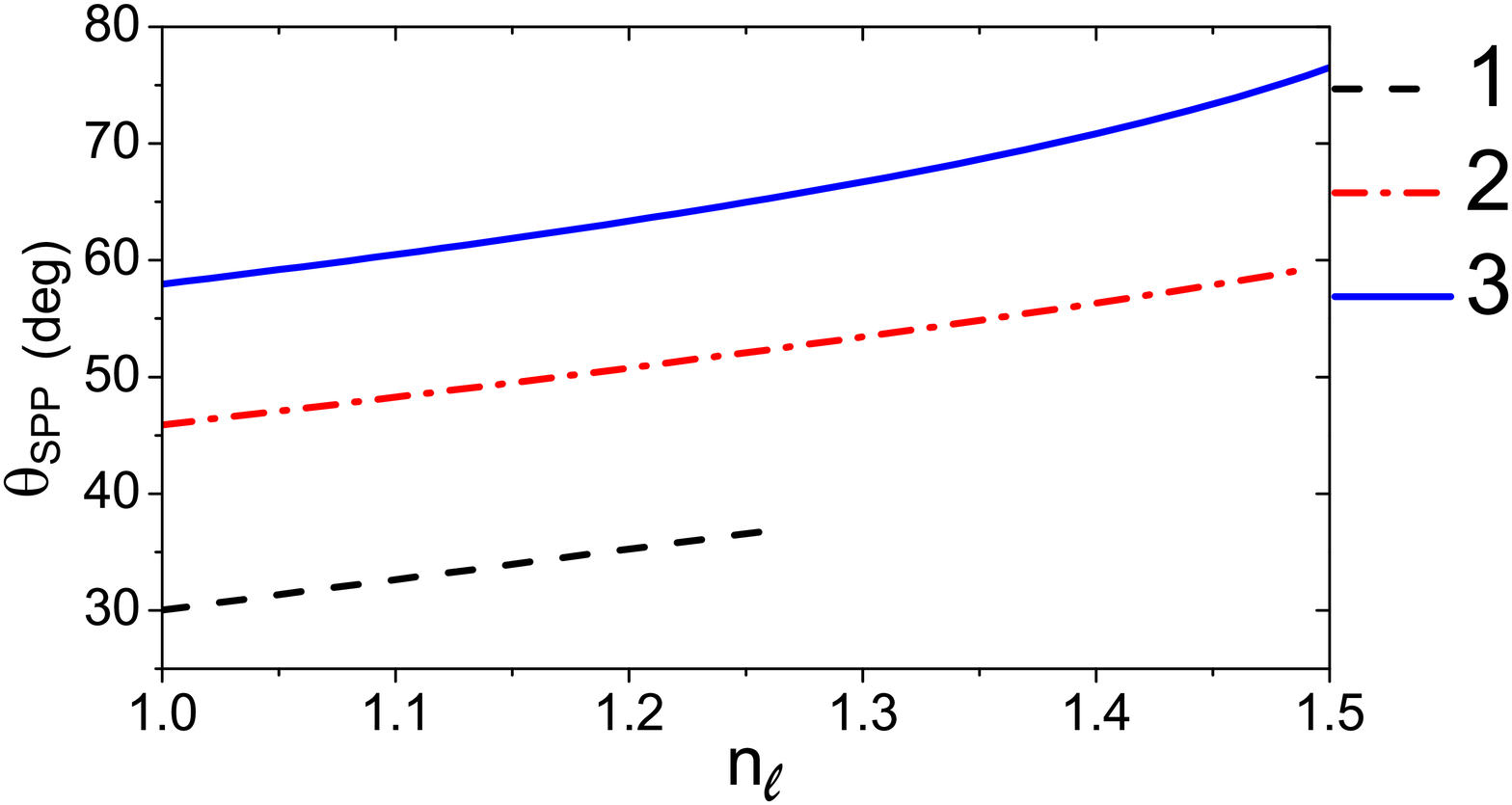}}
\subfigure[]{\includegraphics[width=0.45\textwidth]{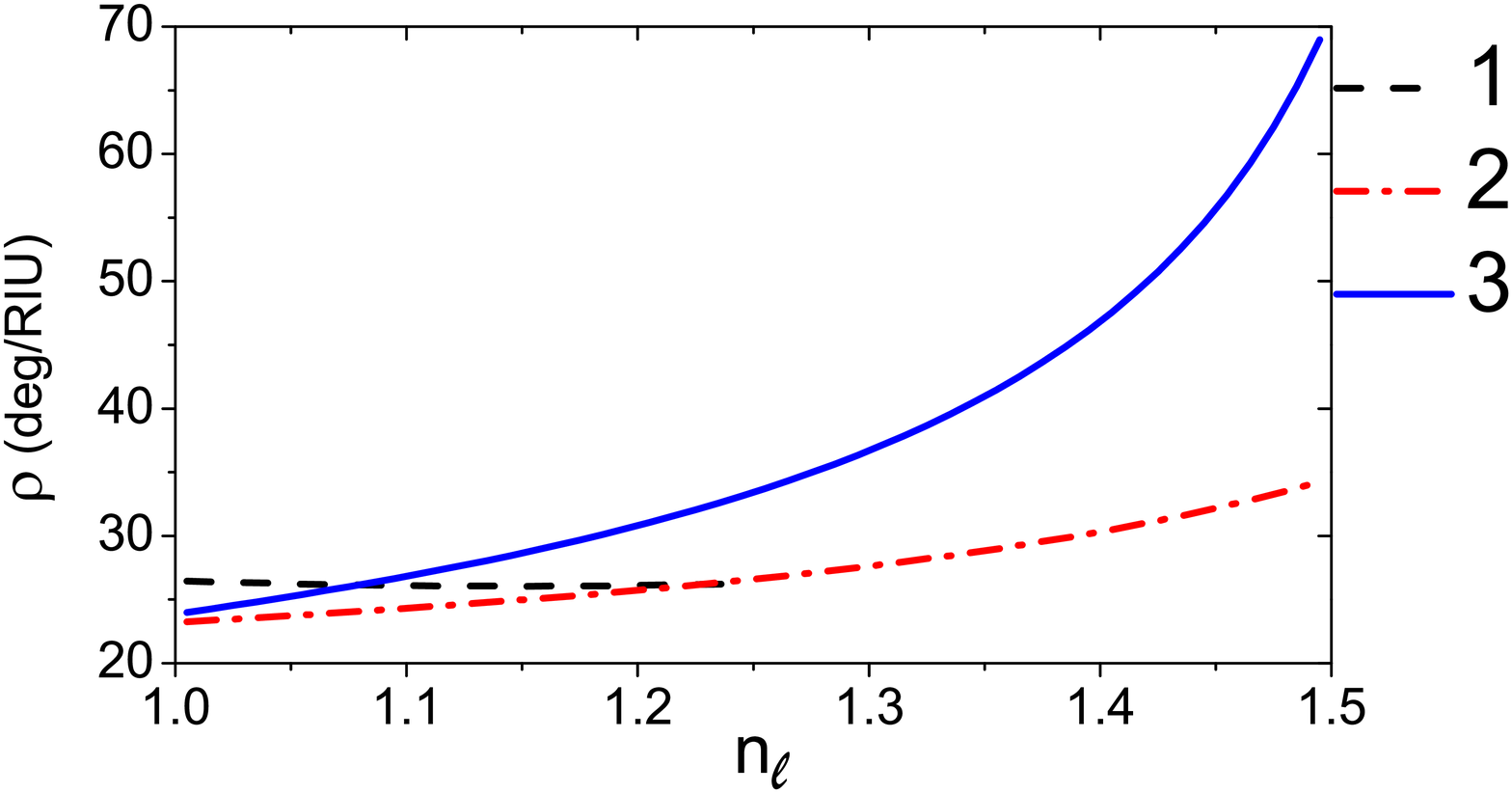}}
\caption{Same as Fig.~\ref{thetarho1} except that the volume fraction of silver nanoparticles is $\fnp=0.01$. }\label{thetarho2}
\end{center}
\end{figure}

Figure~\ref{thetarho2} provides the predicted values of $\thetaSPP$ and $\rho$ when $\fnp=0.01$. As indicated
by Eq.~(\ref{predict}) and confirmed
by a comparison of Figs.~\ref{thetarho1} and \ref{thetarho2}, SPP waves should be excited by less obliquely
incident light when the silver nanoparticles are present. However, the maximum $\rho$ increases from
about 30~deg/RIU in Fig.~\ref{thetarho1}(b) for $\fnp=0$ to about 69~deg/RIU in Fig.~\ref{thetarho2}(b) for $\fnp=0.01$ for branch~3.

\begin{figure}[!ht]
\begin{center}
\subfigure[]{\includegraphics[width=0.45\textwidth]{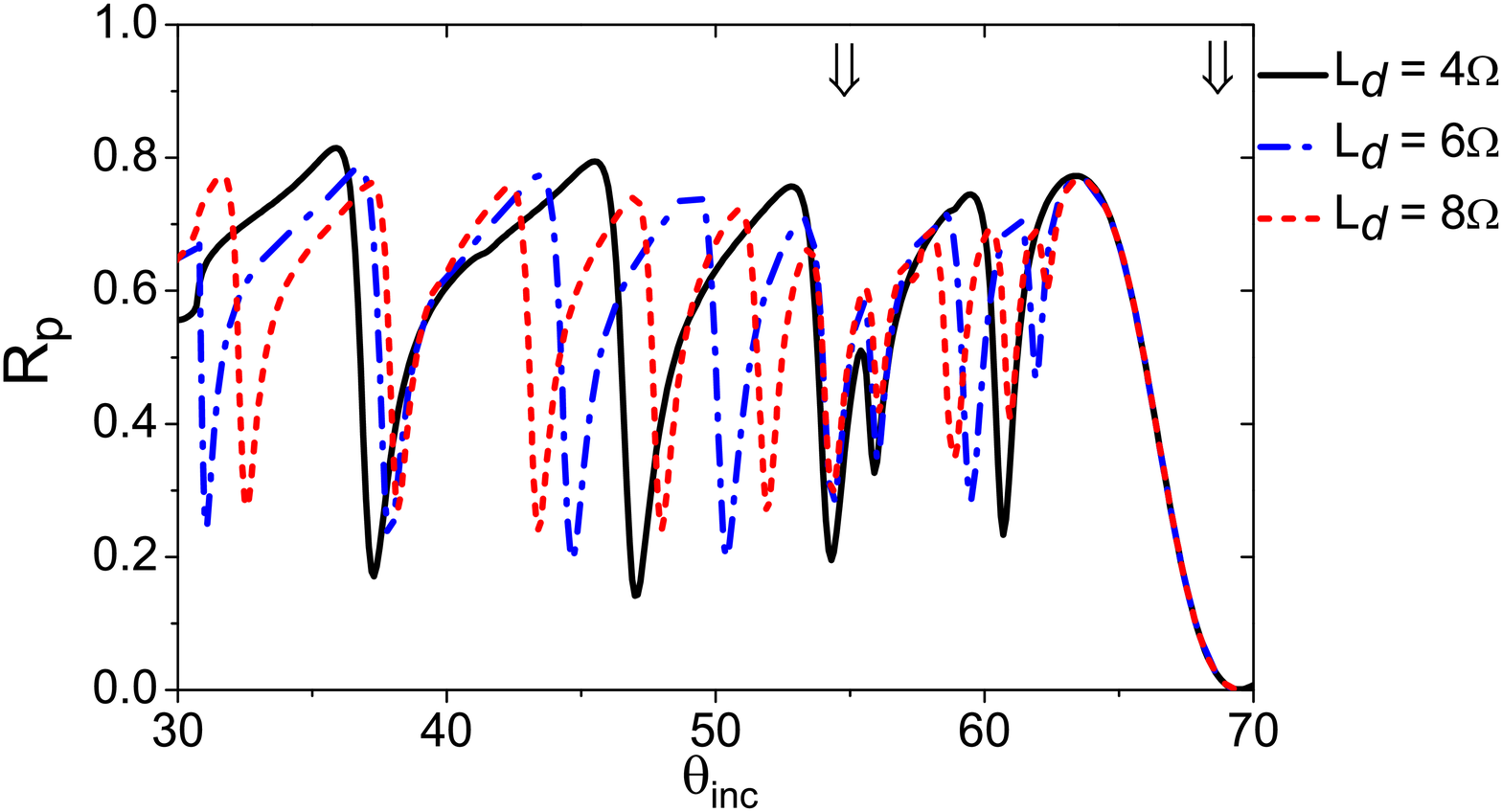}}
\subfigure[]{\includegraphics[width=0.45\textwidth]{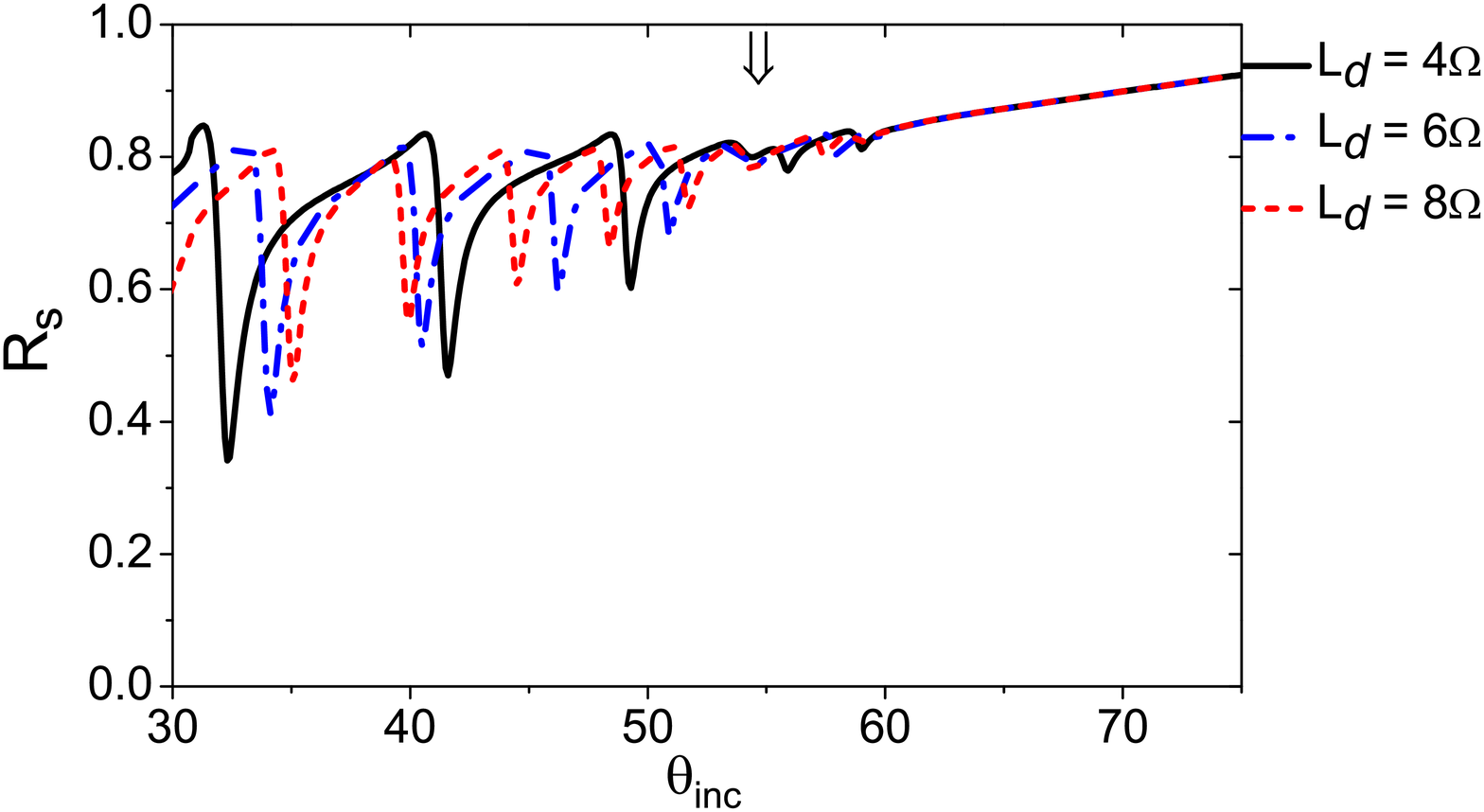}}
\caption{Same as Fig.~\ref{Rps_1} except that the volume fraction of silver nanoparticles is $\fnp=0.01$. }\label{Rps_2}
\end{center}
\end{figure}

When $\nfl=1.33$, Fig.~\ref{thetarho2}(a) indicates that SPP waves can be excited
at  $\thetaSPP\in\left\{54.3^\circ,67.9^\circ\right\}$ in the prism-coupled configuration. Thus, the number of SPP waves predicted reduces by unity when 1\% volume of chiral STF is
occupied by the silver nanoparticles. This reduction is confirmed by the plots of $\Rp$ versus $\thetainc$ in Fig.~\ref{Rps_2}(a),
the minimums of $\Rp$ indicating SPP-wave excitation being located at
$\thetainc\in\left\{54.3^\circ,68.4^\circ\right\}$. Moreover, incident $s$-polarized light can excite an SPP wave at $\thetainc=54.3^\circ$,
according to Fig.~\ref{Rps_2}(b).

\begin{figure}[!ht]
\begin{center}
\includegraphics[width=0.45\textwidth]{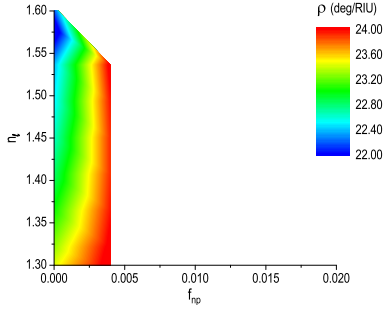}
\caption{The dynamic  sensitivity $\rho$ as  a function of the
refractive index $\nfl$ of the fluid and the volume fraction
$\fnp$ of silver nanoparticles for solutions on branch~1 in Figs.~\ref{kappa1} and \ref{kappa2}.}\label{rho_1}
\end{center}
\end{figure}

\begin{figure}[!ht]
\begin{center}
\includegraphics[width=0.45\textwidth]{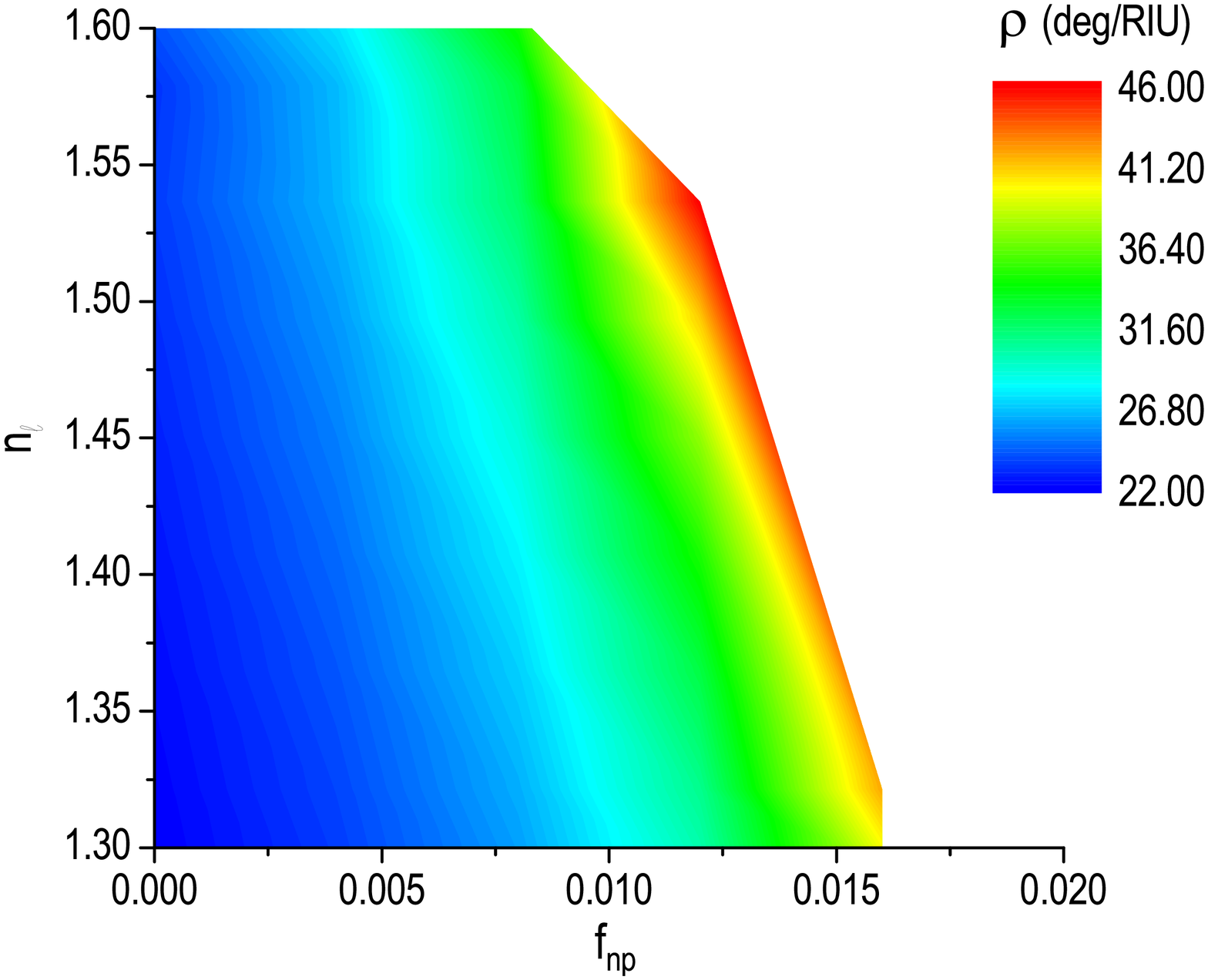}
\caption{Same as Fig.~\ref{rho_1} except for branch 2. }\label{rho_2}
\end{center}
\end{figure}

\begin{figure}[!ht]
\begin{center}
\includegraphics[width=0.45\textwidth]{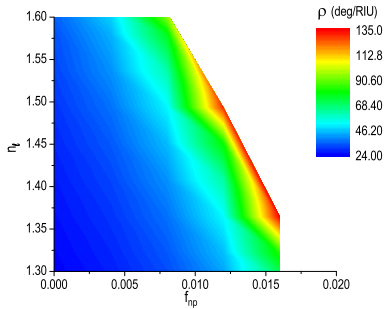}
\caption{Same as Fig.~\ref{rho_1} except for branch 3. }\label{rho_3}
\end{center}
\end{figure}

To find the sweet spot for the highest but practical value of the dynamic
sensitivity $\rho$, we have mapped it as a function of $\fnp$ and $\nfl$ in Figs. \ref{rho_1}-\ref{rho_3} for the three branches of the solutions. We chose $\nfl\in[1.3,1.6]$, since most liquids have refractive index in this range. 

Figure \ref{rho_1} shows that, on branch 1, SPP waves exist only when 
$\fnp\leq 0.004$ to detect liquids with $\nfl\in(1.3,1.6)$ and the change in $\rho$ is 
very small. For the chosen prism and the chiral STF, the SPP waves on this branch 
are excited with  $\thetaSPP\in \left[\sim33^\circ, \sim41^\circ\right]$.
Figure \ref{rho_2} shows that on branch~2, (i) $\rho$ can increase from about 
$22$~deg/RIU to $46$~deg/RIU and (ii) the detection over the entire chosen range of 
$\nfl$ is possible only with $\fnp\leq0.008$ with a maximum $\rho$ of about 
$35$~deg/RIU. Furthermore, the SPP waves on  branch~2 are excited with 
$\thetaSPP\in\left[\sim49^\circ,\sim65^\circ\right]$. Finally, Fig. \ref{rho_3}  that on 
branch~3, (i) $\rho$ can increase up to $135$ deg/RIU, and (ii) the detection 
over  the full chosen range of $\nfl$ is possible only with $\fnp\leq0.008$ 
with a maximum $\rho$ of about $98$ deg/RIU. The SPP waves on branch 3 are excited 
with $\thetaSPP\in\left[\sim65^\circ,\sim85^\circ\right]$.

\subsection{Chiral STF with absorbing dielectric nanoparticles}
\begin{figure}[!ht]
\begin{center}
\includegraphics[width=0.45\textwidth]{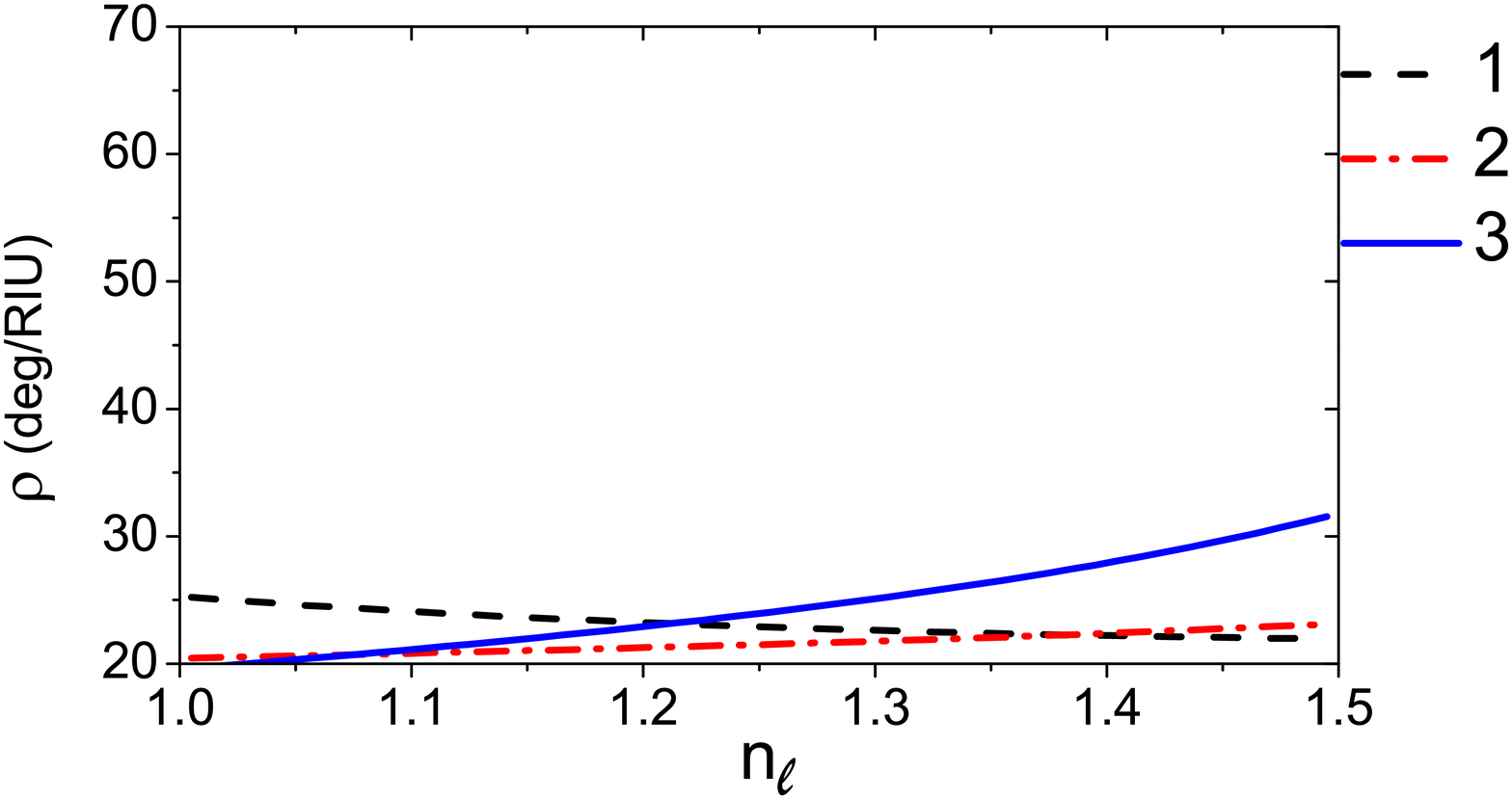}
\caption{Same as Fig.~\ref{thetarho2}(b), except that silver nanoparticles of relative permittivity $-18.2947+0.04808i$ are replaced by absorbing
dielectric nanoparticles  of relative permittivity $18.2947+0.4808i$.
}\label{dielectric}
\end{center}
\end{figure}

To see if the enhancement of dynamic sensitivity is due to absorption loss induced by silver nanoparticles in the chiral STF or if it is due to the negative real part of the permittivity of the silver nanoparticles, we computed $\rho$ when the silver nanoparticles of relative permittivity $(0.0562+4.2776i)^2=-18.2947+0.04808i$ are replaced by absorbing
dielectric nanoparticles  of relative permittivity $(4.2776+0.0562i)^2=18.2947+0.4808i$.
 The data for $\fnp=0.01$ are shown in Fig.~\ref{dielectric}. A comparison of this figure with 
Figs.~\ref{thetarho1}(b) and  \ref{thetarho2}(b) shows that the dynamic  sensitivity is unaffected by the deployment of the absorbing dielectric nanoparticles---in contrast to the deployment of the same volume fraction of silver nanoparticles.
 
 \subsection{Chiral STF with gold nanoparticles}
 Also, we performed  calculations with gold nanoparticles ($\nnp=0.1834+3.4332i$) deployed instead of silver nanoparticles, while $\fnp=0.005$ fixed and  $\nfl\in[1,1.5]$ variable. Branch 3 disappeared.
 The enhancement of  dynamic  sensitivity was negligible with gold nanoparticles, with the maximum value of $\rho$ being $24$ deg/RIU for branch 1 and $26$ deg/RIU for branch 2. These values of maximum $\rho$ are much lower than those obtained with 0.5\% volume fraction of silver nanoparticles:
$54$ deg/RIU for branch 1, $29$ deg/RIU for branch 2, and $24$ deg/RIU for branch 3.
  
 \section{Concluding remarks}\label{con}
The multiple-SPP-waves-based sensing  of a fluid uniformly  infiltrating a chiral STF impregnated with metal nanoparticles  was studied theoretically, when the chiral STF is deployed in the prism-coupled configuration \cite{Homola2003}. Comparison was made with the case  when the nanoparticles are absent \cite{ML2012}. The Bruggemann homogenization formalism was used in two different ways to model the three principal relative permittivity scalars  of the metal-nanoparticle-impregnated chiral STF flooded by the fluid.

The main conclusion was that the dynamic sensitivity of the SPP-wave-based optical sensor can increase 
significantly when the chiral STF is impregnated with silver nanoparticles, provided that the
silver volume fraction does not exceed about 1\%. The enhancement in the dynamic sensitivity is not due to the absorption introduced by the silver nanoparticles, and should be attributed to the  local-field enhancement due to the metal nanoparticles \cite{Kalele} and the possible interaction of particle plasmon-polaritons and surface plasmon-polaritons \cite{KSL,LIAL}. Gold nanoparticles did not enhance dynamic sensitivity as well as silver nanoparticles.


\begin{thebibliography}{99}%

\bibitem{Homola2003} 
Homola J (2003) 
Present and future of surface plasmon resonance biosensors.
Anal. Bioanal. Chem. 377:528--539

\bibitem{AZL}
Abdulhalim I,   Zourob M,    Lakhtakia A (2008)
Surface plasmon resonance for biosensing:
A mini-review. Electromagnetics 28:214--242 

\bibitem{CZM}
 Couture C, Zhao S S,   Masson J-F (2013)
 Modern surface plasmon resonance for bioanalytics
and biophysics.
Phys. Chem. Chem. Phys. 15:11190--11216 

\bibitem{M2007} 
Maier SA (2007)
Plasmonics: Fundamentals and Applications. 
Springer, New York

\bibitem{LMG}
Luong JHT, Male KB, Glennon JD (2008)
Biosensor technology: Technology push versus market pull.
Biotechnol. Adv. 26:492--500

\bibitem{EH2001} 
Hutter E, Cha S, Liu J-F, Park J, Yi J, Fendler JH, Roy D (2001) 
Role of substrate metal in gold nanoparticle enhanced surface plasmon resonance.
J. Phys. Chem. 105:8--12 

\bibitem{HWL}
 Hao P,   Wu Y,     Li F (2011)
 Improved sensitivity of wavelength-modulated surface
plasmon resonance biosensor using gold nanorods.
 Appl. Opt. 50:5555--5558

\bibitem{Bedford}
Bedford EE,  Spadavecchia J,  Pradier C-M,   Gu FX (2012)
Surface plasmon resonance biosensors incorporating gold nanoparticles.
Macromol. Biosci. 12:724--739

\bibitem{BHbook}
 Bohren CF, Huffman DR (1983) Absorption and Scattering of Light by Small Particles. Wiley, New York  
 
 \bibitem{Kalele}
Kalele SA, Tiwari NR, Gosavi SW, Kulkarni SK (2007)
Plasmon-assisted photonics at the nanoscale.
J. Nanophoton. 1:012501

\bibitem{KSL}
Lee K-S, El-Sayed MA (2006)
Gold and silver nanoparticles in sensing and imaging: Sensitivity of plasmon response to size, shape, and metal composition.
J. Phys. Chem. 110:19220--19225

\bibitem{LIAL}
Li A, Isaacs S, Abdulhalim I, Li S (2015)
Ultrahigh enhancement of electromagnetic fields by exciting
localized with extended surface plasmons.
J. Phys. Chem. C 119:19382--19389

\bibitem{Saha}
 Saha K,  Agasti SS,   Kim C,   Li X,  Rotello VM (2012)
Gold nanoparticles in chemical and biological sensing.
Chem. Rev. 112:2739--2779

\bibitem{SPL2013} 
Swiontek SE, Pulsifer DP, Lakhtakia A (2013) 
Optical sensing of analytes in aqueous solutions with a multiple surface-plasmon-polariton-wave platform.
Sci. Rep. 3:1409 

\bibitem{PML2013} 
Polo JA Jr, Mackay TG, Lakhtakia A (2013)
Electromagnetic Surface Waves: A Modern Perspective. 
Elsevier, Waltham, MA

\bibitem{ML2012} 
Mackay TG, Lakhtakia A (2012) 
Modeling chiral sculptured thin films as platforms for surface-plasmonic-polaritonic optical sensing.
IEEE. Sens. J. 12: 273--280 

\bibitem{FA2015} 
Abbas F, Naqvi QA, Faryad M (2015) 
Multiplasmonic optical sensor using sculptured nematic thin films.
Plasmonics.  DOI: 0.1007/s11468-015-9920-7


%
\bibitem{LM2005} 
Lakhtakia A, Messier R (2005) 
Sculptured Thin Films: Nanoengineered Morphology and Optics. 
SPIE Press, Bellingham, WA

\bibitem{AL2007} 
Lakhtakia A (2007) Toward optical sensing of metal nanoparticles using chiral sculptured thin films.
J. Nanophoton. 1:019502

\bibitem{HWH1998} 
Hodgkinson I, Wu Qh, Hazel J (1998) 
Empirical equations for the principal refractive indices and column angle of obliquely deposited films of tantalum oxide, titanium oxide, and zirconium oxide.
Appl. Opt. 37:2653--2659

\bibitem{Mackay2007}
Mackay TG (2007) On the effective permittivity of silver-insulator nanocomposites. J.
Nanophoton. 1:019501 



\end{thebibliography}
\end{document}